\documentclass[traditabstract]{aa}

\usepackage{graphicx}
\usepackage{txfonts}
\usepackage{array}
\usepackage{amssymb}
\usepackage{natbib}
\usepackage{lscape}
\usepackage{rotating}
\usepackage{longtable}
\newcommand{\hd}{HD\,166734}
\newcommand{\kms}{km\,s$^{-1}$}
\newcommand{\xmm}{{\sc{XMM}}\emph{-Newton}}
\newcommand{\swift}{{\em Swift}}
\begin{document}

\title{An X-ray view of \hd, a massive supergiant system\thanks{Based on observations collected with {\it{Swift}} and the ESA science mission \xmm , an ESA Science Mission with instruments and contributions directly funded by ESA Member States and the USA (NASA).}}

\author{Ya\"el~Naz\'e\thanks{F.R.S.-FNRS Research Associate.}
\and Eric~Gosset\thanks{F.R.S.-FNRS Senior Research Associate.}
\and Laurent Mahy\thanks{F.R.S.-FNRS Postdoctoral Researcher.}
\and Elliot Ross Parkin
}

\institute{Groupe d'Astrophysique des Hautes Energies, STAR, Universit\'e de Li\`ege, Quartier Agora (B5c, Institut d'Astrophysique et de G\'eophysique), All\'ee du 6 Ao\^ut 19c, B-4000 Sart Tilman, Li\`ege, Belgium\\
\email{naze@astro.ulg.ac.be}
}

\authorrunning{Naz\'e et al.}
\titlerunning{X-rays from \hd }
\abstract{The X-ray emission of the O+O binary \hd\ was monitored using \swift\ and \xmm\ observatories, leading to the discovery of phase-locked variations. The presence of an $f$ line in the He-like triplets further supports a wind-wind collision as the main source of the X-rays in \hd. While temperature and absorption do not vary significantly along the orbit, the X-ray emission strength varies by one order of magnitude, with a long minimum state ($\Delta(\phi)\sim0.1$) occurring after a steep decrease. The flux at minimum is compatible with the intrinsic emission of the O-stars in the system, suggesting a possible disappearance of colliding wind emission. While this minimum cannot be explained by eclipse or occultation effects, a shock collapse may occur at periastron in view of the wind properties. Afterwards, the recovery is long, with an X-ray flux proportional to the separation $d$ (in hard band) or to $d^2$ (in soft band). This is incompatible with an adiabatic nature for the collision (which would instead lead to $F_X\propto 1/d$), but could be reconciled with a radiative character of the collision, though predicted temperatures are lower and more variable than in observations. An increase in flux around $\phi\sim0.65$ and the global asymmetry of the light curve remain unexplained, however. }
\maketitle

\section{Introduction}
Stellar winds are a key ingredient in the lives of massive stars, but their exact properties are still a subject of debate. To gain further insight into these winds, there are several observational opportunities, such as analyses of P\,Cygni profiles in the UV spectra of single stars or of colliding wind emission in binaries. Indeed, when two massive stars form a gravitationally bound pair, their winds collide, and the characteristics of this interaction depends on the relative strengths of the winds. 

These collisions produce signatures over a wide range of wavelengths; in particular, for some massive binaries, the post-shock temperature is so high that X-rays are emitted. The main observational characteristics of such an emission is its variability due to a variety of factors (for a review, see \citealt{rau16}): the absorption towards the collision zone changes as it is seen through different wind densities over an orbital period, the collision strength depends on the stellar separation in eccentric binaries, and occultation of the collision zone by the stellar bodies may occur, as well as hysteresis effects. The observed variations thus depend on stellar and orbital parameters, but also on the orientation of the system with respect to Earth and on the details of plasma physics. 

In this context, it would be extremely interesting to compare systems with similar orbits and orientations but different wind momentum ratios as they would provide a kind of ``controlled experiment'' to test our understanding of such wind-wind collisions. Recently, we have analysed the case of the eccentric binary WR21a \citep{gos16}. Its period is $\sim$32\,d and its eccentricity $e\sim0.7$, which are close to those of another massive binary: \hd. However, these binaries have very different wind momentum ratios: $\dot M_1 v_{\infty,1}/\dot M_2 v_{\infty,2}$ is about 7 for WR21a \citep{gos16}, but only $\sim$3 for \hd\ \citep[hereafter Paper I]{gos17}. Since \hd\ has never been studied at X-ray wavelengths, we have undertaken a specific X-ray monitoring of this system, with the hope of constraining the properties of its wind-wind collision.

This paper investigates the X-ray emission of \hd\ as observed by \swift\ and \xmm. Section 2 provides some information on the target, Sect. 3 presents the data and their reduction, Sect. 4 explains the analysis,  Sect. 5 discusses the results, and Sect. 6 summarizes and concludes.

\section{The target}
The binarity of \hd\ was first reported by \citet{wol63} but the monitoring of \citet{con80} provided the first orbital solution of the system. These authors found the period to be 34.54\,d and the masses of the stars to be similar despite different spectral types (O7.5If+O9I).  
In view of their masses, \citet{con80} also proposed that eclipses  occur in \hd, and eclipses were indeed reported for the system by \citet{ote05}, although there is only one per orbit. 
No additional visual companion was detected around the central binary \citep{mas09,san14}. 

Recently, Paper I revisited the system using dedicated photometric and spectroscopic monitorings. The orbital parameters of \hd\ were then refined and the stellar characteristics of each of its components determined, thanks to the derivation of the individual spectra using disentangling methods and the spectral fitting made with atmosphere models. Table \ref{param} summarizes the new parameters of the system, which we will use in this follow-up paper.

\begin{table}
\centering
\caption{Stellar and orbital parameters derived in Paper I.}
\label{param}
\begin{tabular}{lcc}
\hline\hline
Parameter & Primary & Secondary\\
\hline
spectral types & O7.5If & O9I(f) \\
$T_{\rm eff}$\,[kK] & 32.0$\pm$1.0 & 30.5$\pm$1.0\\
$\dot M$\,[$10^{-6}$M$_{\odot}$\,yr$^{-1}$]& 9.07& 3.02\\
$v_{\infty}$\,[\kms] & 1386 & 1331 \\
$\log(L_{BOL}/L_{\odot})$ & 5.840$\pm$0.092 & 5.732$\pm$0.104 \\
\hline
$T_0$ (periastron) & \multicolumn{2}{c}{2\,452\,195.064$\pm$0.036}\\
$P$ [d] & \multicolumn{2}{c}{34.537723$\pm$0.001330}\\
$e$ & \multicolumn{2}{c}{0.618$\pm$0.005}\\
$i\,[^{\circ}]$ & \multicolumn{2}{c}{63.0$\pm$2.7}\\
$\omega\,[^{\circ}]$ & \multicolumn{2}{c}{236.183$\pm$0.786} \\
$a\,\sin(i)$\,[R$_{\odot}$] & 76.23$\pm$0.89 & 89.85$\pm$1.05\\
$M$\,[M$_{\odot}$] & 39.5$\pm$5.4 & 33.5$\pm$4.6\\
$R_{mean}$\,[R$_{\odot}$]& 27.5$\pm$2.3 & 26.8$\pm$2.4\\
\hline
\end{tabular}
\end{table}

\section{Observations and data reduction} 

\begin{table*}
\centering
\footnotesize
\caption{Journal of the X-ray observations:  X for \xmm\ observations and S for \swift\ exposures. HJD correspond to dates at mid-exposure, and the corresponding phases were calculated using the ephemeris of Paper I; exposure times correspond to on-axis values (for pn if \xmm). The \xmm\ count rates correspond to the sum of MOS1, MOS2, and pn values. We note that the \swift\ XID notation follows the ObsID naming scheme (there is thus no S-1 as 00034304001 has zero seconds of useful time). }
\label{journal}
\begin{tabular}{lccccc}
\hline\hline
XID & ObsID (exp. time) & HJD & $\phi$ & \multicolumn{2}{c}{Count Rates (cts\,s$^{-1}$)}\\
    &                   &     &        &  0.5--1.5\,keV & 1.5--10.0\,keV \\
\hline
X-1 & 0500030101 (47.6\,ks) & 2454531.604 & 0.65&  1.707$\pm$0.007   &  0.927$\pm$ 0.005  \\ 
X-2 & 0790180601  (8.0\,ks) & 2457480.796 & 0.04&  0.131$\pm$ 0.005  &  0.054$\pm$ 0.004  \\ 
S-2 & 00034304002 (0.7\,ks) & 2457428.242 & 0.52& 0.0743$\pm$ 0.0145 & 0.0395$\pm$ 0.0100 \\
S-3 & 00034304003 (3.2\,ks) & 2457429.903 & 0.57& 0.0592$\pm$ 0.0050 & 0.0403$\pm$ 0.0040 \\
S-4 & 00034304004 (3.4\,ks) & 2457433.340 & 0.67& 0.0727$\pm$ 0.0052 & 0.0508$\pm$ 0.0043 \\
S-5 & 00034304005 (2.3\,ks) & 2457436.002 & 0.75& 0.0674$\pm$ 0.0059 & 0.0467$\pm$ 0.0049 \\
S-6 & 00034304006 (1.0\,ks) & 2457438.692 & 0.82& 0.0863$\pm$ 0.0126 & 0.0462$\pm$ 0.0085 \\
S-7 & 00034304007 (3.7\,ks) & 2457441.646 & 0.91& 0.0334$\pm$ 0.0034 & 0.0251$\pm$ 0.0030 \\
S-8 & 00034304008 (2.8\,ks) & 2457444.605 & 0.99& 0.0068$\pm$ 0.0018 & 0.0040$\pm$ 0.0014 \\
S-9 & 00034304009 (3.8\,ks) & 2457447.656 & 0.08& 0.0070$\pm$ 0.0015 & 0.0035$\pm$ 0.0011 \\
S-10& 00034304010 (2.8\,ks) & 2457453.179 & 0.24& 0.0249$\pm$ 0.0033 & 0.0228$\pm$ 0.0031 \\
S-11& 00034304011 (2.0\,ks) & 2457455.908 & 0.32& 0.0360$\pm$ 0.0056 & 0.0331$\pm$ 0.0052 \\
S-12& 00034304012 (1.4\,ks) & 2457456.626 & 0.34& 0.0484$\pm$ 0.0064 & 0.0429$\pm$ 0.0061 \\
S-13& 00034304013 (1.4\,ks) & 2457476.611 & 0.92& 0.0266$\pm$ 0.0050 & 0.0293$\pm$ 0.0054 \\
S-14& 00034304014 (1.9\,ks) & 2457480.297 & 0.03& 0.0066$\pm$ 0.0020 & 0.0036$\pm$ 0.0016 \\
S-15& 00034304015 (1.5\,ks) & 2457484.685 & 0.16& 0.0136$\pm$ 0.0032 & 0.0158$\pm$ 0.0035 \\
S-16& 00034304016 (0.8\,ks) & 2457486.085 & 0.20& 0.0254$\pm$ 0.0066 & 0.0164$\pm$ 0.0053 \\
S-17& 00034304017 (0.3\,ks) & 2457486.887 & 0.22& 0.0176$\pm$ 0.0089 & 0.0087$\pm$ 0.0063 \\
S-18& 00034304018 (3.8\,ks) & 2457494.506 & 0.44& 0.0518$\pm$ 0.0040 & 0.0349$\pm$ 0.0033 \\
\hline
\end{tabular}
\end{table*}

\subsection{\xmm }
The first \xmm\ observation of \hd, available in the archives, was taken in March 2008 (Principal Investigator Valencic). It shows a bright source, with a count rate compatible with {\it ROSAT} values. However, in view of the variations seen in the \swift\ data (see below), we requested in March 2016 a second short exposure to sample the phase of minimum flux. It was granted under Director Discretionary Time. Both datasets were reduced with SAS (Science Analysis Software) v15.0.0 using calibration files available in Spring 2016 and following the recommendations of the \xmm\ team\footnote{SAS threads, see \\ http://xmm.esac.esa.int/sas/current/documentation/threads/ }. 

The EPIC (European Photon Imaging Camera) observations, taken in the full-frame mode and with the medium filter (to reject optical/UV light), were filtered to keep only the best-quality data ({\sc{pattern}} of 0--12 for MOS and 0--4 for pn). Background flares were detected in the first observation, and they were cut before engaging in further analyses. A source detection was performed on each EPIC dataset using the task {\it edetect\_chain} on the 0.5--1.5 (soft) and 1.5--10.0 (hard) energy bands and for a log-likelihood of 10. This task searches for sources using a sliding box and determines the final source parameters from point spread function (PSF) fitting; the final count rates correspond to equivalent on-axis, full PSF count rates (Table~\ref{journal}).  

We then extracted EPIC spectra of \hd\ using the task {\it{especget}} in circular regions of 36\arcsec\ radius (to avoid nearby sources) centred on the best-fit positions found for each observation. For the background, a circular region of 40\arcsec\ radius was chosen in a region devoid of sources and as close as possible to the target; its relative position with respect to the target is the same for both observations. Dedicated ARF (Ancillary Response File) and RMF (Redistribution Matrix File) response matrices, which are used to calibrate the flux and energy axes, respectively, were also calculated by this task. EPIC spectra were grouped with {\it{specgroup}} to obtain an oversampling factor of five and to ensure that a minimum signal-to-noise ratio of 3 (i.e.\ a minimum of 10 counts) was reached in each spectral bin of the background-corrected spectra. 

EPIC light curves of \hd\ were extracted for time bins of 100\,s, 500\,s, and 1\,ks, in the same regions as the spectra, and in the same energy bands (plus the total one, 0.5--10.\,keV band) as the source detection. They were further processed by the task {\it epiclccorr}, which corrects for loss of photons due to  vignetting, off-axis angle, or other problems such as bad pixels. In addition, to avoid very large errors and bad estimates of the count rates, we discarded bins displaying effective exposure time lower than 50\% of the time bin length. Our previous experience with \xmm\ has shown us that including such bins degrades the results. As the background is much fainter than the source, in fact too faint to provide a meaningful analysis, three sets of light curves were produced and analysed individually: the raw source+background light curves, the background-corrected light curves of the source, and the light curves of the sole background region. The results found for the raw and background-corrected light curves of the source are indistinguishable. 

\begin{figure*}
\includegraphics[width=8.5cm]{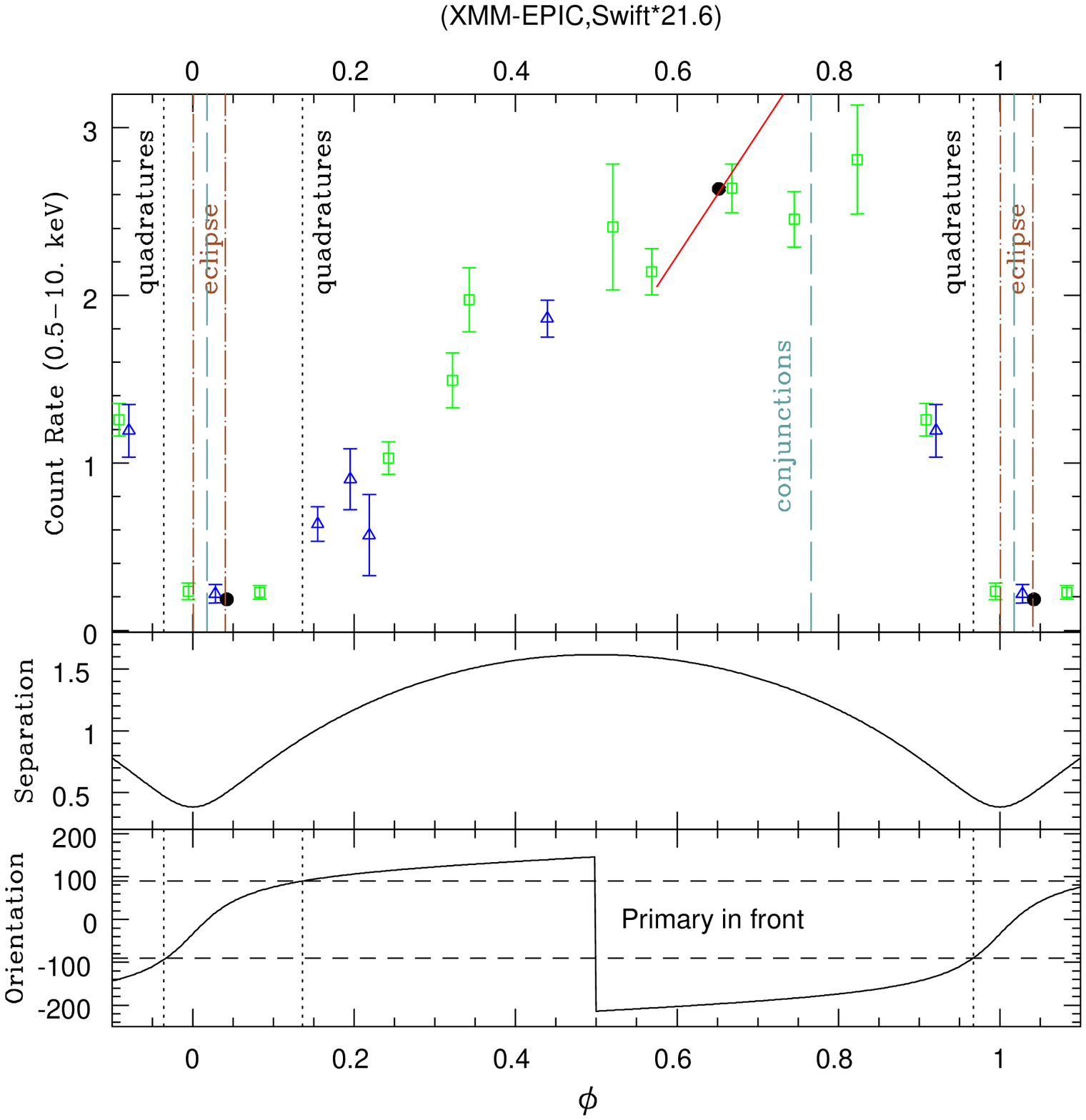}
\includegraphics[width=8.5cm]{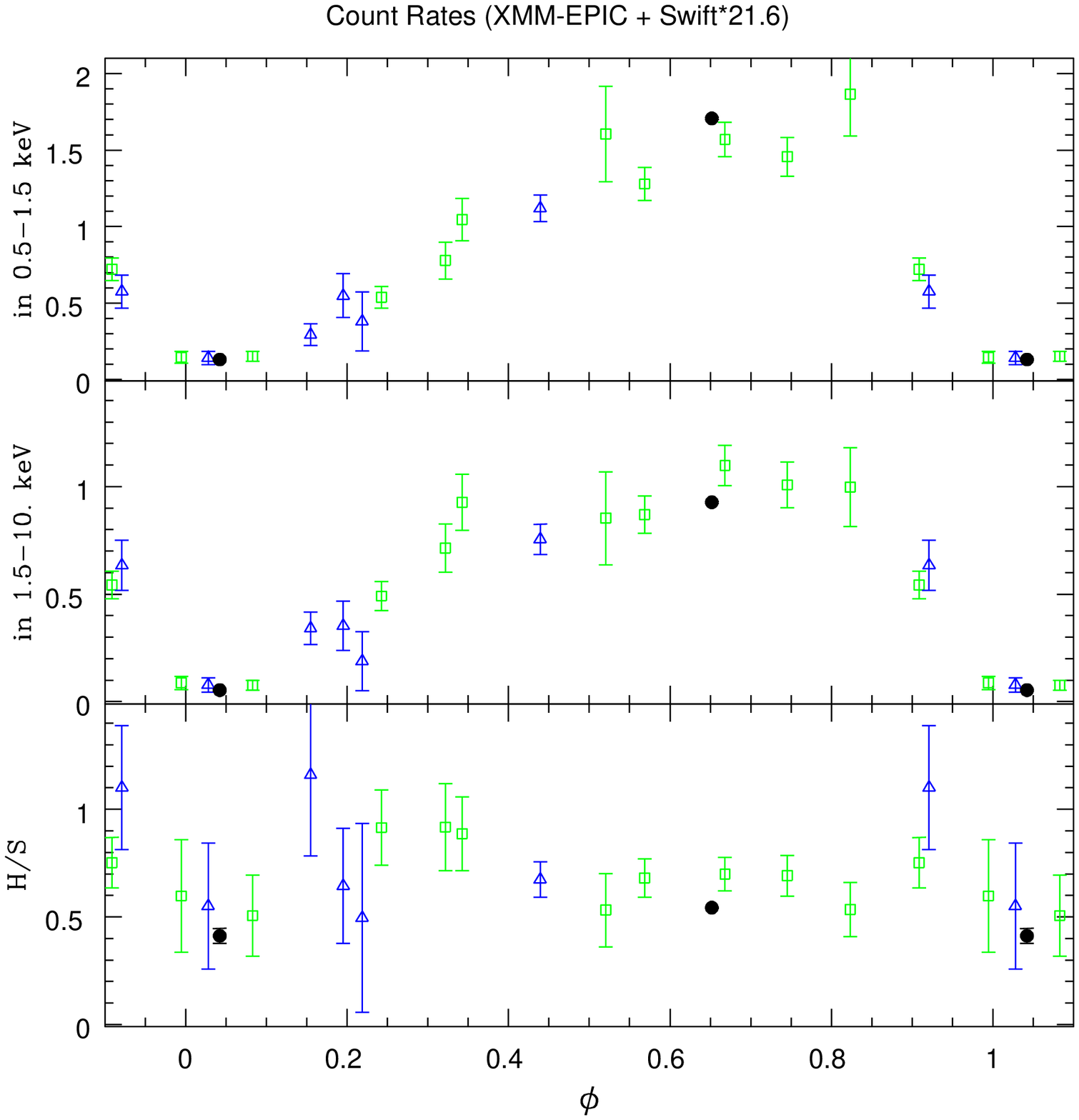}
\caption{Variation in the count rates and hardness ratio with orbital phase for \xmm\ (black dots) and {\it{Swift}} data (empty green squares for S-2 to 12 from Feb-March 2016, blue triangles for S-13 to 18 from April 2016; see Table \ref{journal}). For all energy bands, {\it{Swift}} count rates and their errors were multiplied by 21.6 to clarify the trends (no further treatment was made for adjustment). In the top left  panel, the straight red line represents the trend detected in the intra-pointing  light curve of X-1. To help the comparison with physical parameters, the bottom left panels provide the orbital separation (in units of the semi-major axis $a$) as well as a position angle defined as zero when the O7.5I star (primary) is in front and 180$\degr$ when the O9I star (secondary) is in front. The vertical lines in the top left  panel indicate the phases of the quadratures, conjunctions, and optical eclipse start/end. All data are phased according to the ephemeris of Paper I.}
\label{lc}
\end{figure*}

The EPIC count rates of \hd\ in the first observation (1.6\,cts\,s$^{-1}$ for pn and 0.5\,cts\,s$^{-1}$ for MOS) are close to the pile-up limits ($\sim$2 and $\sim$0.5\,cts\,s$^{-1}$ for pn and MOS, respectively). We therefore checked the presence of pile-up using the task {\it epatplot}, which did not show any clear evidence of pile-up. Furthermore, we extracted spectra of \hd\ in an annulus centred on the source and considering only single events ({\sc{pattern}}=0). Again, no evidence for pile-up is detected as spectral fitting provide results in agreement with those found using spectra derived in the usual way. 

Only RGS (Reflection Grating Spectrometer) spectra of the first observation have enough counts for a scientific analysis. No flare filtering was applied, however, as the background  light curve showed an elevated level throughout the exposure rather than localized flares. The source and background spectra were extracted in the default regions as \hd\ has no neighbour of similar X-ray brightness. Dedicated response files were calculated for both orders and both RGS, and were subsequently attached to the source spectra for analysis.

\subsection{{\it{Swift}}}
At our request, \hd\ was observed 17 times by {\it{Swift}} between February and April 2016 (Table~\ref{journal}). These data were retrieved from the HEASARC (High-Energy Astrophysics Science Archive Research Center). Since the target is quite bright ($V$=8.42), UVOT (Ultraviolet/Optical Telescope) cannot be used and XRT (X-ray Telescope) data, which were taken in PC (Photon Counting) mode, could suffer from optical loading, requiring a check.  Using the dedicated web tool\footnote{http://www.swift.ac.uk/analysis/xrt/optical\_tool.php} for the appropriate properties of our target ($T_{\rm eff}$=30.5--32\,kK, $BC$ between --2.8 and --3.0), we found that spurious events from optical loading would at most lead to $10^{-5}$\,cts\,s$^{-1}$. This is well below the detected values for the target, hence the optical loading can be considered  negligible. 

XRT data taken in PC mode were processed locally using the XRT pipeline of HEASOFT (High-Energy Astrophysics Software) v6.18 with calibrations available in Spring 2016. Corrected count rates in the same energy bands as \xmm\ were obtained for each observation from the UK online tool\footnote{http://www.swift.ac.uk/user\_objects/} (Table~\ref{journal}). It also provided the best-fit position for the full dataset,
which is similar to Simbad's value. The Simbad position was thus used to extract the source spectra within Xselect in a circular region of 50\arcsec\ radius (as recommended). They were binned using {\it grppha} in a similar manner to that used for the \xmm\ spectra. Following the recommendations of the {\it{Swift}} team, the largest possible background region was chosen, i.e.\ an annulus of outer radius 100\arcsec. The most recent RMF matrix from the calibration database was used while specific ARF response matrices were calculated for each dataset using {\it xrtmkarf}, considering the associated exposure map. 

Because of the small number of counts and  the excellent repeatability with phase of the light curve (see next section), we combined some \swift\ spectra taken at similar phases using the {\sc ftools} {\it addpha} and {\it addarf}: S-8, S-9, and S-14 ($\phi\sim0.03$); S-10, S-15, and S-16 ($\phi\sim0.20$); S-11 and S-12 ($\phi\sim0.33$); S-7 and S-13 ($\phi\sim0.91$). The weights for ARF combinations were in proportion to the number of counts in the individual spectra. In contrast, the spectra of datasets S-2, S-6, and S-17 had too few counts to be useful, so we discarded them from the spectral analyses.

\section{Results}

\subsection{Light curves}
At first glance, the data directly show large variations in the X-ray brightness of \hd\ (see Table \ref{journal} and left panel of Fig. \ref{lc}). The amplitude of the variations is very large: the ratio between the extreme count rates exceeds one order of magnitude. To understand the nature of such changes, it is important to note that the \swift\ data were taken during two consecutive orbits of the system. Folding them with the orbital period yields  a remarkable agreement, demonstrating the repeatability of these changes (Fig. \ref{lc}). In particular, it is worth noting the quasi-identity  between two data points taken in different orbits (S-7 and S-13): as both belong to the steep descending branch, any phase shift would have been readily detected if the wrong timescale had been used. Moreover, the two \xmm\ datasets, taken 8 years apart (corresponding to 85 orbits), also agree well with the \swift\ light curve. Clearly, the variations of the X-ray emission of \hd\ are phase-locked in nature. Fortunately, the \swift\ data cover the full 34.54\,d orbit in a homogeneous manner. 

The light curve of \hd\ shows a flat minimum lasting $\Delta(\phi)$=0.1 around periastron (from $\phi$=0.99 to 0.08). This minimum is preceded by a steep, order-of-magnitude decrease, beginning after $\phi\sim0.8$. It is followed by a shallower increase up to $\phi\sim0.5$ and a quasi-flattening afterwards. The same behaviour is detected in both the soft and hard bands, so that there are no drastic changes in hardness (see right panel of Fig. \ref{lc}): in the more sensitive \xmm\ data, the ratio between hard and soft count rates only changes from 0.41$\pm$0.03 (at minimum flux) to 0.543$\pm$0.004, i.e. there is a significant (4$\sigma$) but limited softening at minimum flux; in the \swift\ data, there is also a slight increase in hardness near $\phi$=0.2--0.3, but it remains within 2$\sigma$ of the other values. 

\begin{figure}
\includegraphics[width=8.5cm]{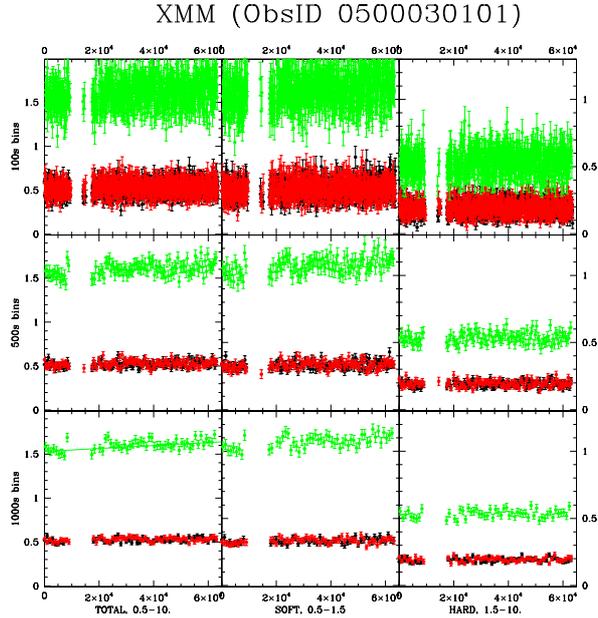}
\caption{Intra-pointing light curves of the first \xmm\ observations (pn in green, MOS1 in black, MOS2 in red). The top, middle, and bottom panels show the light curves with 100\,s, 500\,s, and 1000\,s time bins, respectively, while the left, central, and right panels display the total, soft, and hard light curves. The ordinates in cts\,s$^{-1}$ are given on the left side for the total light curves, but on the right side for the soft and hard light curves. In the bottom left panel, the best-fit linear trend is superimposed onto the pn data.}
\label{lcxmm}
\end{figure}

The first \xmm\ observation covers half a day. We have thus analysed the intra-pointing variability of that dataset (Fig. \ref{lcxmm}) using $\chi^2$ tests for three different null hypotheses (constancy, linear variation, quadratic variation). The improvement in the $\chi^2$ when increasing the number of parameters in the model (e.g. linear trend vs constancy) was also determined by means of Snedecor F tests \citep[nested models; see Sect.~12.2.5 in][]{lind76}. While all light curves are formally compatible with a constant level (even for significance levels of 1\%), we found that the soft and total pn light curves were significantly better fitted by a linear trend. This trend is also detected in soft and total MOS light curves, but at the 5\% significance level only (which can be explained by the lower sensitivity of these cameras). The slope is 2.5$\times10^{-6}$\,cts\,s$^{-2}$ for the sum of all EPIC data. When overplotted on the global light curve (left panel of Fig. \ref{lc}), this increasing trend corresponds to the increase observed between the S-3 and S-4 datasets. It is, however, steeper than the overall shallow increase seen for $\phi$=0.5--0.8. 

\subsection{Global spectral fitting}

We fitted the spectra under Xspec v12.9.0i, considering reference solar abundances from \citet{asp09}. As X-ray lines are clearly detected on both low- and high-resolution spectra, absorbed optically thin thermal emission models (of the type $wabs\times phabs \times \sum apec$) were considered. The first absorption component represents the interstellar contribution. It was fixed to $8.1\times10^{21}$\,cm$^{-2}$, calculated using the reddening of the target ($E(B-V)=1.33$, from $B-V$=1.07 and $(B-V)_0=-0.26$), and the conversion factor from \citet{gud12}. 

\begin{figure*}
\includegraphics[width=6cm]{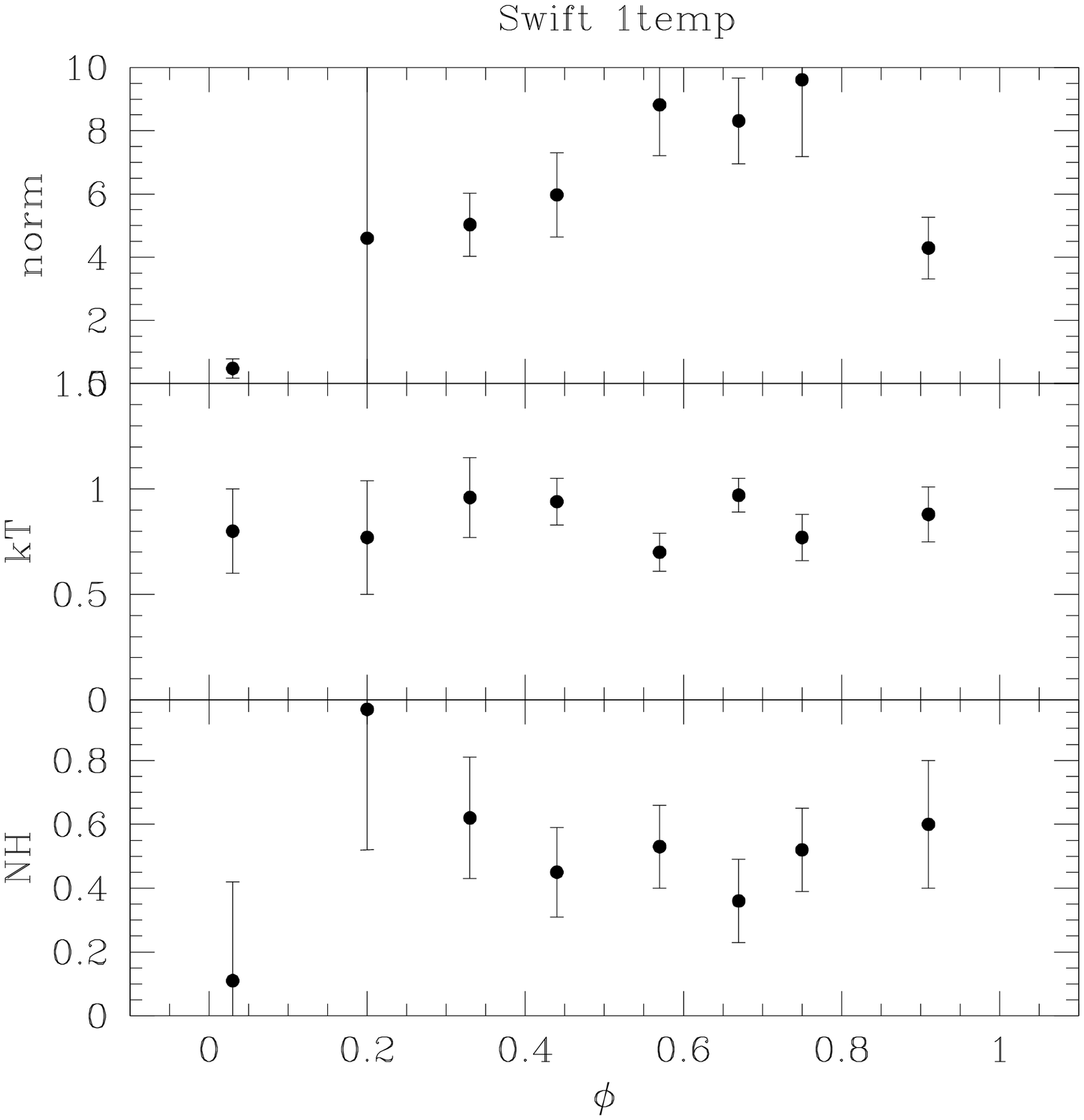}
\includegraphics[width=6cm]{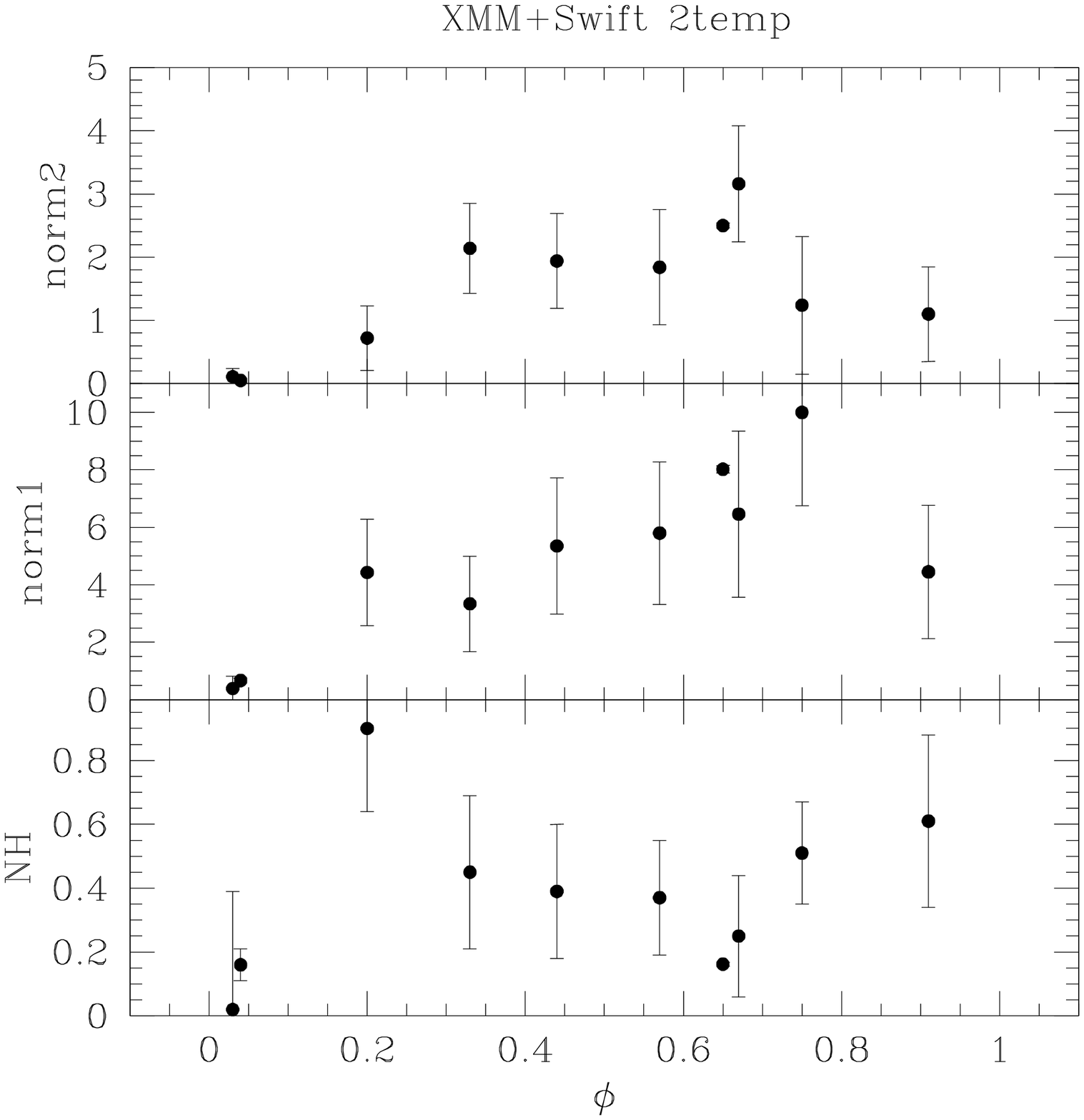}
\includegraphics[width=6cm]{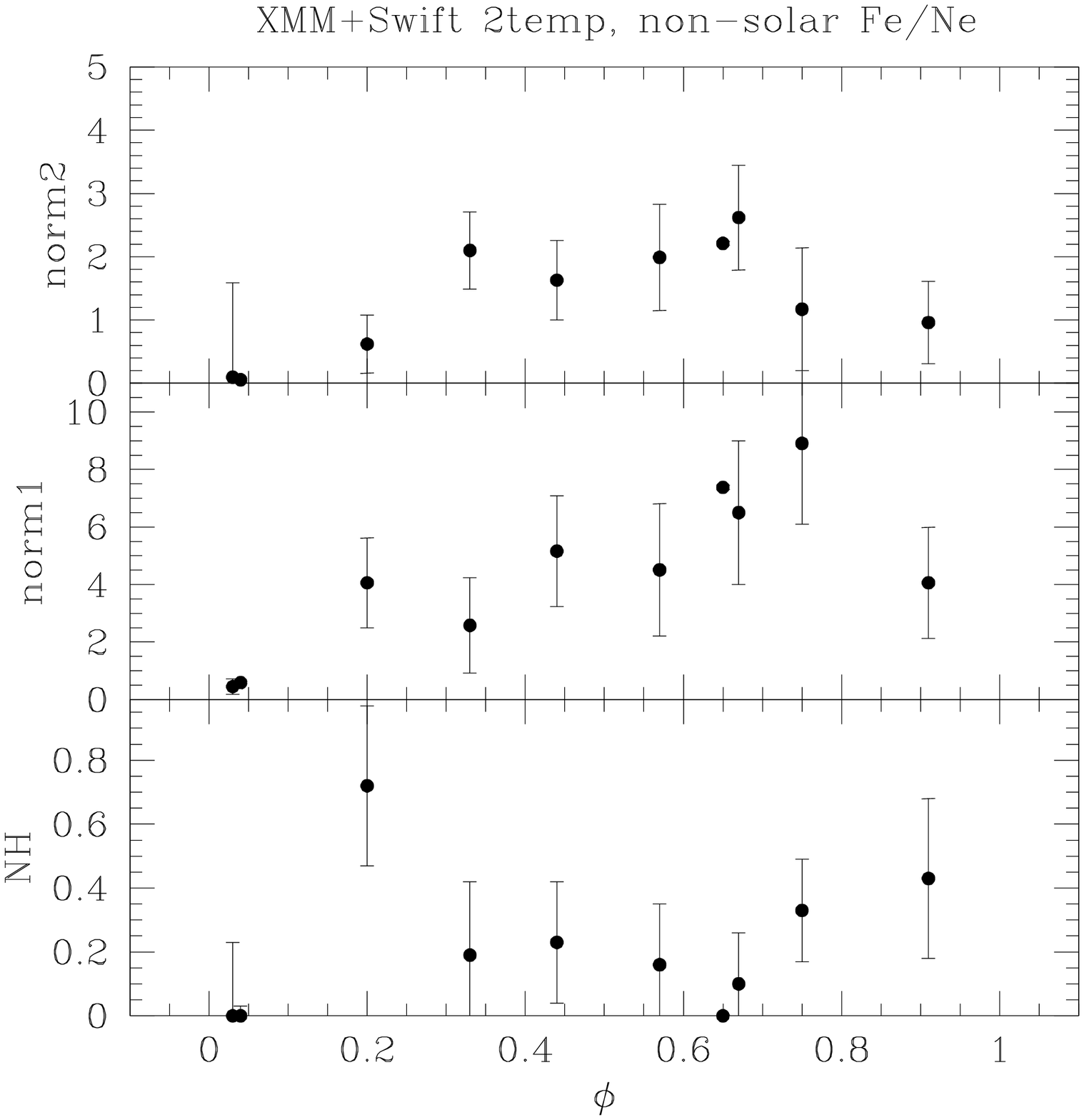}
\includegraphics[width=6cm]{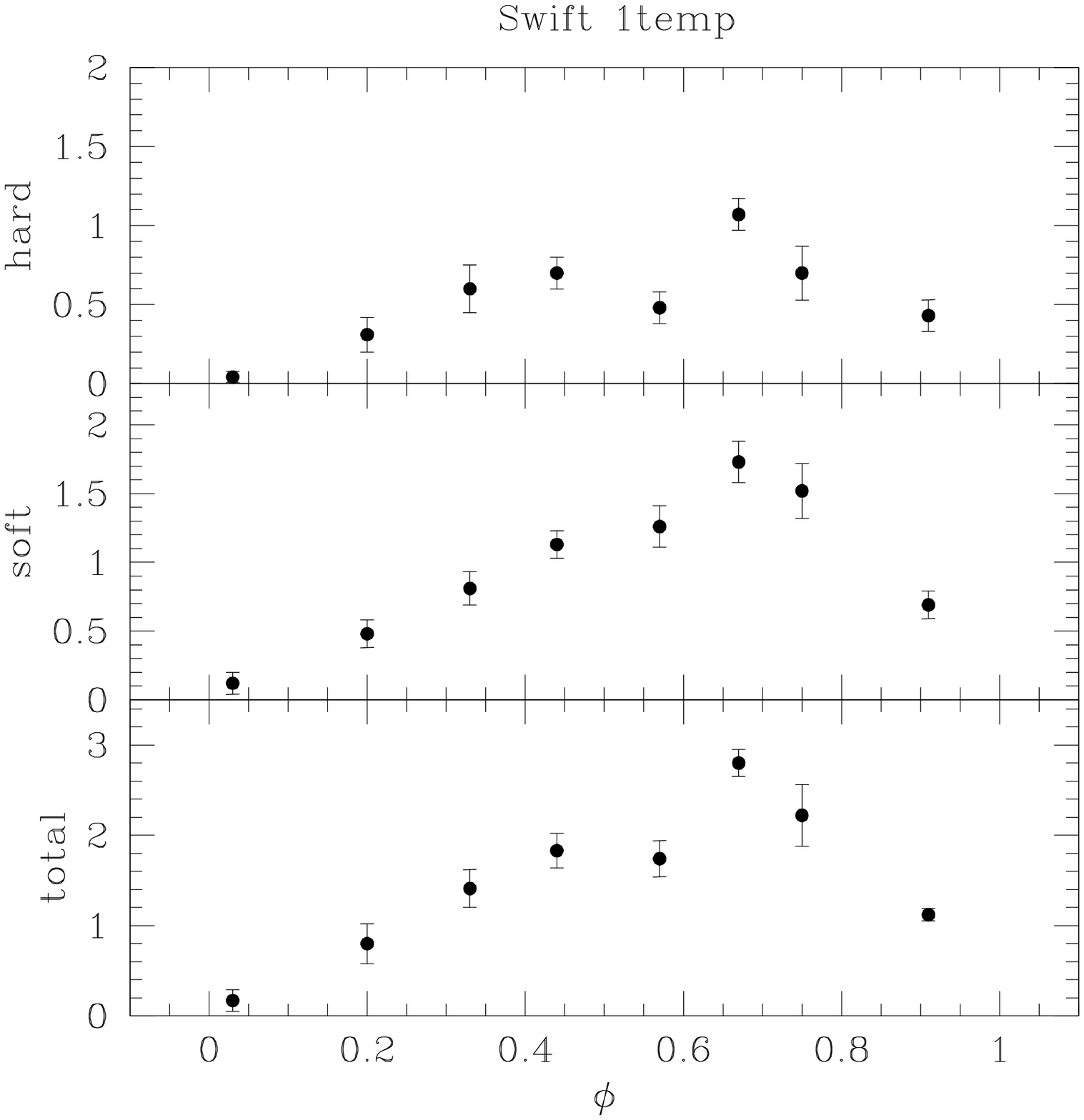}
\includegraphics[width=6cm]{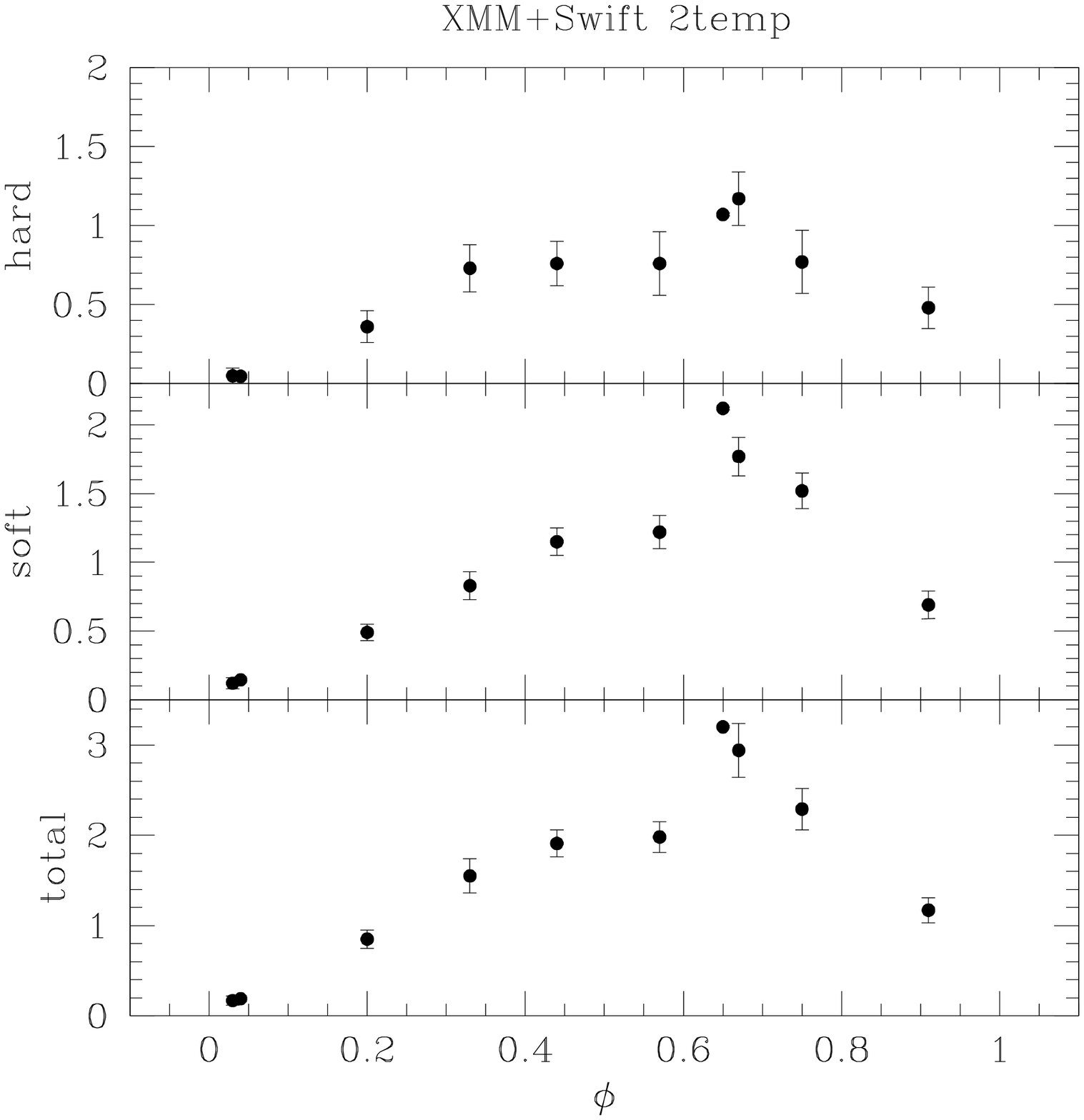}
\includegraphics[width=6cm]{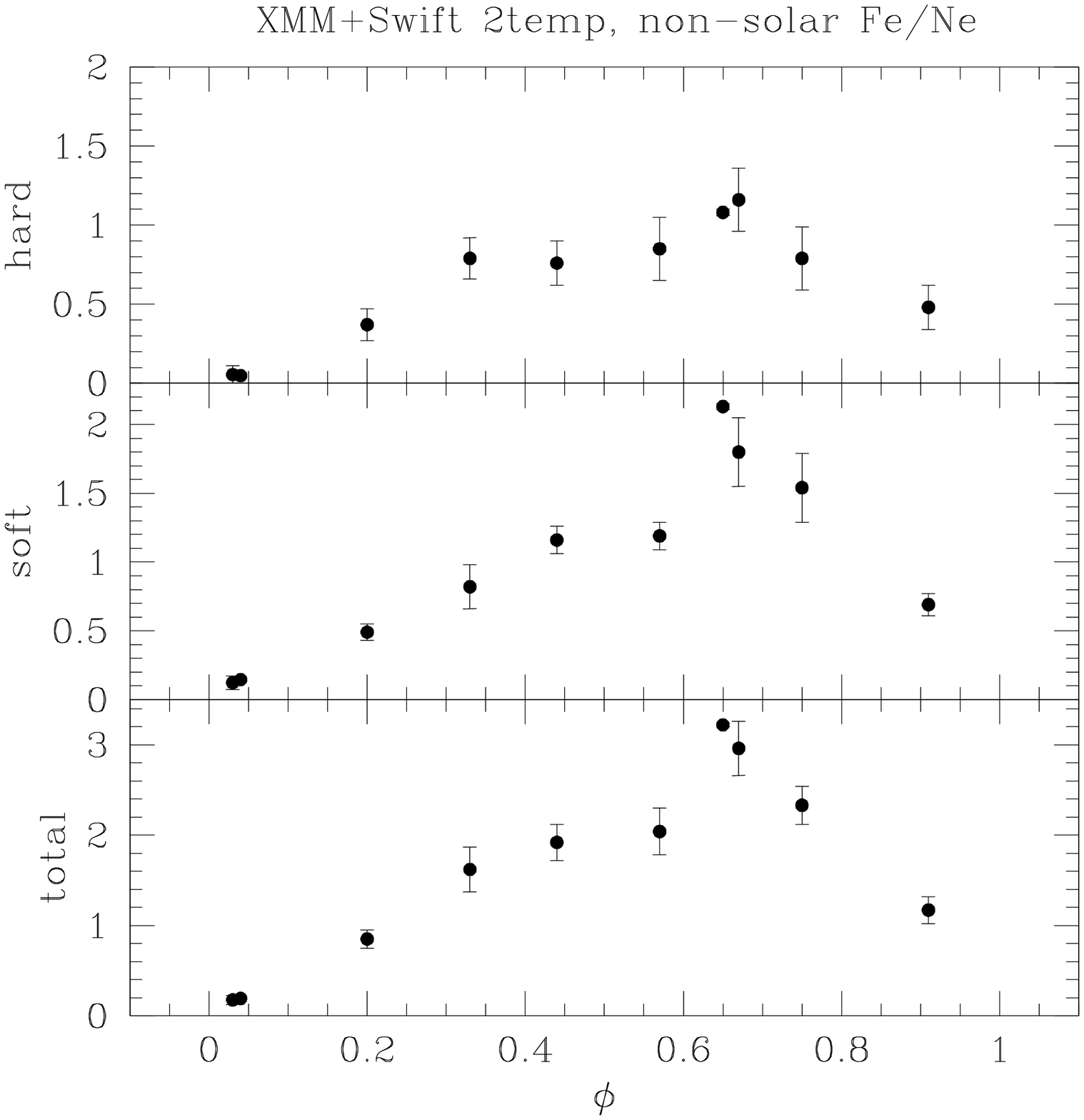}
\caption{Results of the spectral fits for the three types of fits performed (Table \ref{fits}). The three panels on the top line provide the variations with phase of the spectral parameters ($norm$ in units 10$^{-3}$\,cm$^{-5}$, $kT$ in keV, and $N_{\rm H}$ in units 10$^{22}$\,cm$^{-2}$) while the three panels on the bottom line yield observed fluxes in units 10$^{-12}$\,erg\,cm$^{-2}$\,s$^{-1}$. The panel pairs in the left, middle, and right columns correspond to results from single temperature fits, two-temperature fits with solar abundances, and two-temperature fits with non-solar abundances, respectively.}
\label{fitparam}
\end{figure*}

The \swift\ spectra have a low number of counts, they can thus be fitted with a single thermal component (Table \ref{fits}, Fig. \ref{fitparam}). The normalization factors and the fluxes follow the light curve based on count rates (see previous subsection and Fig. \ref{lc}). The derived temperature appears constant at 0.85\,keV on average. The absorption appears slightly lower at minimum flux, slightly higher just afterwards, and very stable at other phases; however,  these changes are at the 1--2$\sigma$ level only. Fits forcing the absorption to remain at a constant value (the mean of the values obtained earlier) results in a marked (though not large) increase in the $\chi^2$ for the spectra taken at $\phi\sim0.15$. 

All \xmm\ spectra were fitted simultaneously, i.e. all EPIC+RGS for X-1 and all EPIC for X-2. These spectra could not be fitted with a single temperature, but the sum of two thermal components provides an excellent fit except in the 1.--2.\,keV range for the EPIC spectra of X-1. For this dataset, using individual absorptions for each thermal component does not improve the quality of the fit but the $\chi^2$ can be improved by changing some abundances ({\it not} the global metallicity). Since the RGS spectra clearly show Ne, Mg, and Fe lines (see Sect. 4.3) and since these elements are responsible for lines in the 1.--2.\,keV range, we only allow the abundances of these elements to vary\footnote{Paper I found departures from solar abundances for HeCNO, but the abundances of the primary and secondary are different and the X-ray data cannot be used to constrain these abundances for the hot plasma, hence we kept these elements to solar values. However, as demonstrated below for Ne and Fe, changing abundances would not affect our conclusions.}. In all the trials we performed, the Fe abundance became subsolar and the Mg abundance stayed close to 1. The Ne abundance has a mixed behaviour. It appears slightly subsolar when fitted on RGS data alone, but is clearly oversolar when fitted on EPIC or EPIC+RGS spectra. Physically, this result is puzzling, but we know from experience that abundances found in EPIC spectra are not to be trusted blindly and that they have limited informative value. We therefore performed two sets of fits, one with solar abundances and a second set with Ne and Fe abundances fixed to the best-fit values found by fitting the full first \xmm\ dataset (i.e. both EPIC and RGS data). Since we detect no clear change in temperature between the two \xmm\ datasets, and since there was no significant temperature variation in the \swift\ data, the temperatures were also fixed to the best-fit values found for the first \xmm\ dataset. Results for both cases (solar and non-solar abundances) are provided in Table \ref{fits} and Fig. \ref{fitparam}: the trends are similar in both cases. Furthermore, they agree well with the results from simple fits to the \swift\ spectra. It should be further noted that no significant change in absorption is detected between the two \xmm\ datasets and that the average temperature varies by only 20\% between the two \xmm\ datasets (confirming the hardness ratio results). Finally, the hard flux appears stable in the $\phi=0.3-0.8$ interval except for a peak at $\phi\sim0.65$ which corresponds well to the increase detected in the first \xmm\ dataset. 

\begin{table*}
\centering
\caption{Results of the global spectral fits.  }
\label{fits}
\begin{tabular}{lccccccc}
\hline\hline
\multicolumn{8}{l}{Model $wabs_{ism}*phabs*apec$}\\
ID  & $\phi$ & $N_{\rm H}$ & $kT$ & $norm$ & $\chi^2$ (dof) & $F^{\rm obs}_{\rm X}$ & $F^{\rm unabs}_{\rm X}$\\
    &        & (10$^{22}$\,cm$^{-2}$) & (keV) & ($10^{-3}$\,cm$^{-5}$) & & \multicolumn{2}{c}{($10^{-12}$\,erg\,cm$^{-2}$\,s$^{-1}$)}\\
\hline
S-3        & 0.57 & 0.53$\pm$0.13 & 0.70$\pm$0.09 & 8.82$\pm$1.61 & 1.19(19) & 1.74$\pm$0.20 & 6.38  \\
S-4        & 0.67 & 0.36$\pm$0.13 & 0.97$\pm$0.08 & 8.31$\pm$1.36 & 1.02(27) & 2.80$\pm$0.15 & 8.76  \\
S-5        & 0.75 & 0.52$\pm$0.13 & 0.77$\pm$0.11 & 9.61$\pm$2.43 & 1.00(17) & 2.22$\pm$0.34 & 7.57  \\
S-7+13     & 0.91 & 0.60$\pm$0.20 & 0.88$\pm$0.13 & 4.29$\pm$0.98 & 0.4(18)  & 1.12$\pm$0.07 & 3.23  \\
S-8+9+14   & 0.03 & 0.11$\pm$0.31 & 0.80$\pm$0.20 & 0.48$\pm$0.31 & 0.75(5)  & 0.17$\pm$0.12 & 0.81  \\
S-10+15+16 & 0.20 & 0.96$\pm$0.44 & 0.77$\pm$0.27 & 4.60$\pm$7.84 & 1.24(14) & 0.80$\pm$0.22 & 1.97  \\
S-11+12    & 0.33 & 0.62$\pm$0.19 & 0.96$\pm$0.19 & 5.03$\pm$1.00 & 1.84(15) & 1.41$\pm$0.21 & 3.70  \\
S-18       & 0.44 & 0.45$\pm$0.14 & 0.94$\pm$0.11 & 5.97$\pm$1.33 & 0.99(23) & 1.83$\pm$0.19 & 5.58  \\
\hline
\multicolumn{8}{l}{Model $wabs_{ism}*phabs*[apec(0.64keV)+apec(1.52keV)]$ with solar abundances}\\
ID  & $\phi$ & $N_{\rm H}$ & $norm_1$ & $norm_2$ & $\chi^2$ (dof) & $F^{\rm obs}_{\rm X}$ & $F^{\rm unabs}_{\rm X}$ \\
    &        & (10$^{22}$\,cm$^{-2}$) & ($10^{-3}$\,cm$^{-5}$) & ($10^{-3}$\,cm$^{-5}$) &  & \multicolumn{2}{c}{($10^{-12}$\,erg\,cm$^{-2}$\,s$^{-1}$)} \\
\hline
X-1(EP+RGS)& 0.65 &0.162$\pm$0.006& 8.02$\pm$0.13 & 2.50$\pm$0.04 & 1.45(2392) & 3.20$\pm$0.01  & 14.2  \\
X-2(EPIC)  & 0.04 & 0.16$\pm$0.05 & 0.67$\pm$0.08 & 0.05$\pm$0.03 & 1.14(96)   &0.191$\pm$0.008 & 1.02  \\
S-3        & 0.57 & 0.37$\pm$0.18 & 5.80$\pm$2.48 & 1.84$\pm$0.91 & 1.09(19)   & 1.98$\pm$0.17  & 6.86  \\
S-4        & 0.67 & 0.25$\pm$0.19 & 6.46$\pm$2.89 & 3.16$\pm$0.92 & 1.10(27)   & 2.94$\pm$0.30  & 10.8  \\
S-5        & 0.75 & 0.51$\pm$0.16 & 10.0$\pm$3.25 & 1.24$\pm$1.09 & 0.99(17)   & 2.29$\pm$0.23  & 7.99  \\
S-7+13     & 0.91 & 0.61$\pm$0.27 & 4.45$\pm$2.32 & 1.10$\pm$0.75 & 0.61(18)   & 1.17$\pm$0.14  & 3.37  \\
S-8+9+14   & 0.03 & 0.02$\pm$0.37 & 0.39$\pm$0.43 & 0.11$\pm$0.13 & 0.79(5)    & 0.17$\pm$0.05  & 0.96  \\
S-10+15+16 & 0.20 & 0.90$\pm$0.26 & 4.43$\pm$1.85 & 0.72$\pm$0.51 & 1.13(14)   & 0.85$\pm$0.10  & 2.08  \\
S-11+12    & 0.33 & 0.45$\pm$0.24 & 3.34$\pm$1.66 & 2.14$\pm$0.71 & 1.62(15)   & 1.55$\pm$0.19  & 4.37  \\
S-18       & 0.44 & 0.39$\pm$0.21 & 5.35$\pm$2.37 & 1.94$\pm$0.75 & 1.00(23)   & 1.91$\pm$0.15  & 6.41  \\
\hline
\multicolumn{8}{l}{Model $wabs_{ism}*vphabs*[vapec(0.65keV)+vapec(1.88keV)]$ with Ne and Fe abundances set to 1.88 and 0.63 times solar}\\
ID  & $\phi$ & $N_{\rm H}$ & $norm_1$ & $norm_2$ & $\chi^2$ (dof) & $F^{\rm obs}_{\rm X}$ & $F^{\rm unabs}_{\rm X}$ \\
    &        & (10$^{22}$\,cm$^{-2}$) & ($10^{-3}$\,cm$^{-5}$) & ($10^{-3}$\,cm$^{-5}$) &  & \multicolumn{2}{c}{($10^{-12}$\,erg\,cm$^{-2}$\,s$^{-1}$)} \\
\hline
X-1(EP+RGS)& 0.65 &   0.$\pm$0.004& 7.38$\pm$0.07 & 2.21$\pm$0.03 &1.27(2392) & 3.22$\pm$0.02  & 15.2 \\
X-2(EPIC)  & 0.04 &   0.$\pm$0.03 & 0.59$\pm$0.04 & 0.06$\pm$0.02 & 1.12(96)  &0.193$\pm$0.010 & 1.07 \\
S-3        & 0.57 & 0.16$\pm$0.19 & 4.51$\pm$2.30 & 1.99$\pm$0.84 & 1.16(19)  & 2.04$\pm$0.26  & 7.23 \\
S-4        & 0.67 & 0.10$\pm$0.16 & 6.50$\pm$2.49 & 2.62$\pm$0.83 & 1.06(27)  & 2.96$\pm$0.30  & 11.6 \\
S-5        & 0.75 & 0.33$\pm$0.16 & 8.91$\pm$2.81 & 1.17$\pm$0.97 & 0.89(17)  & 2.33$\pm$0.21  & 8.36 \\
S-7+13     & 0.91 & 0.43$\pm$0.25 & 4.06$\pm$1.93 & 0.96$\pm$0.65 & 0.63(18)  & 1.17$\pm$0.15  & 3.50 \\
S-8+9+14   & 0.03 &   0.$\pm$0.23 & 0.45$\pm$0.27 & 0.10$\pm$1.49 & 0.83(5)   & 0.18$\pm$0.05  & 0.89 \\
S-10+15+16 & 0.20 & 0.72$\pm$0.25 & 4.06$\pm$1.56 & 0.62$\pm$0.46 & 1.10(14)  & 0.85$\pm$0.10  & 2.14 \\
S-11+12    & 0.33 & 0.19$\pm$0.23 & 2.58$\pm$1.66 & 2.10$\pm$0.61 & 1.49(15)  & 1.62$\pm$0.25  & 4.87 \\
S-18       & 0.44 & 0.23$\pm$0.19 & 5.16$\pm$1.93 & 1.63$\pm$0.63 & 1.01(23)  & 1.92$\pm$0.20  & 6.72 \\
\hline
\end{tabular}
\\
\tablefoot{The first part of the table presents fits with a single thermal component (\swift\ data only), while the second and third parts list the fitting results for two thermal components of fixed temperatures for solar abundances and for non-solar Ne and Fe abundances. Unabsorbed fluxes are only corrected for the interstellar column ($8.1\times10^{21}$\,cm$^{-2}$). Errors (found using the ``error'' command for the spectral parameters and the ``flux err'' command for the fluxes) correspond to 1$\sigma$; whenever errors were asymmetric, the highest value is provided here. Fluxes are expressed in the 0.5--10.0\,keV band. }
\end{table*}

\begin{figure}
\includegraphics[width=8.5cm]{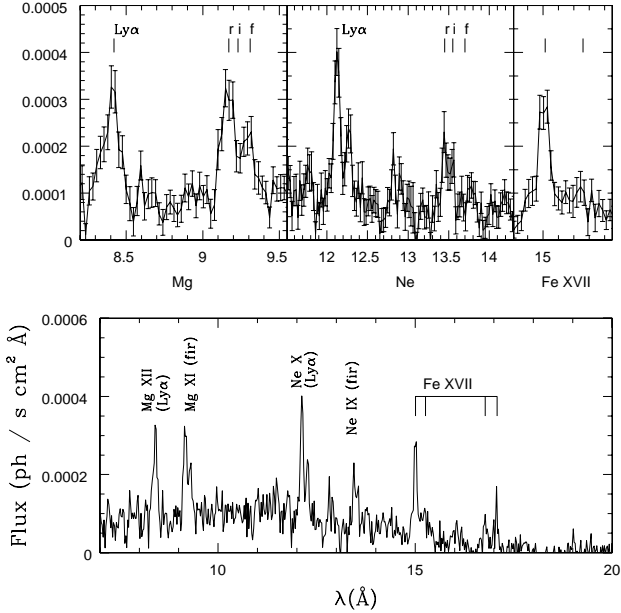}
\caption{The combined RGS spectrum for dataset X-1. The top panel shows a close-up on the strongest lines.}
\label{rgs}
\end{figure}

\subsection{X-ray lines}

Because of the strong interstellar absorption in front of \hd\ and the low sensitivity of RGS at high energies, the combined RGS spectrum (Fig. \ref{rgs}) only shows  a few lines: the Lyman-$\alpha$ and He-like ratios of Ne and Mg, and some Fe\,{\sc xvii} lines near 15\AA\ and 17\AA. Line fitting of the ungrouped raw (i.e. without background subtraction) spectra were performed using the Cash statistics and following two methods. 

First, we used a flat power law (to represent the background level) combined with a sum of Gaussians. For Lyman-$\alpha$ and the Fe doublet at 15\AA, two Gaussians were used with their respective amplitudes fixed to the theoretical value\footnote{ATOMDB, see http://www.atomdb.org/Webguide/webguide.php}. For He-like ``triplets'' (with its $fir$ components), four Gaussians were used and only the relative amplitude of the central two intercombination lines was fixed to the theoretical value. In both cases, the widths and shifts of all Gaussians were assumed to be the same. Table \ref{rgsfit} lists the results of these fits. The most reliable results come from the doublets as they are the strongest and most isolated lines: no significant shift is detected and the line FWHMs are similar to the terminal wind velocity ($\sim$1350\,km\,s$^{-1}$, see Table \ref{param}), as is usual for O-stars \citep{wal07}. The noisier triplets agree with these results, within 2--3$\sigma$. These triplets can be used for two additional derivations. The temperature can be estimated from the $G=(f+i)/r$ ratios, and  from the ratios between the Lyman-$\alpha$ intensities and the combined ($f+i+r$) He-like triplet line strengths; however, the latter ratios must  be corrected for the effect of (interstellar) absorption, as the Lyman-$\alpha$ and He-like triplet lines are at different wavelengths, hence they suffer from different absorption effects. Following the method outlined in \citet{naz12a}, but considering the latest version of ATOMDB, we derive $\log(T)$=6.9--7.0 for Mg and $\log(T)$=6.6--7.0 for Ne. These temperatures agree well with those found in spectral fits (Table \ref{fits} - 0.65, 0.85, and 1.88\,keV corresponding to $\log(T)$=6.9, 7.0, and 7.3, respectively). Furthermore, the $f/i$ line ratio can be used to constrain where the X-ray emitting plasma lies with respect to the star \citep{por01,leu06}. In our case, however, we have the problem that only an upper limit on the $i$ line strength is available, so that only the lower limits of the ratios can be found: $f/i>1.6$ for Mg and $>10$ for Ne, considering the upper limit of the $i$ line strength and the best-fit strength minus $1\sigma$ for the $f$ line. The former value agrees well with the expected theoretical ratio {\it without} any depopulation of the upper level of the $f$ line (1.--2. for temperatures in the $\log(T)$=6.9--7.3 range). The latter value is too high with respect to the theoretical ratio, but considering a 3$\sigma$ uncertainty on the strength of the lines reconciles the observed value with the expected ones. Despite the high uncertainty, it seems clear that the $f$ line exists, and with a strength suggesting no depopulation by UV photons: the X-ray emitting plasma is thus located far from the stellar surfaces. This agrees well with an origin in a colliding-wind shock;  in fact, detecting an $f$ line but no (or faint) $i$ line is usually considered as evidence for the presence of such collisions \citep[e.g.][]{sch04,pol05,sug08,zhe14}.

Second, we used the {\it windprofile} package available for Xspec\footnote{http://heasarc.nasa.gov/xanadu/xspec/models/windprof.html}. It models the line profiles expected for shocks distributed throughout the wind \citep[and references therein]{leu06}. In addition to the line strength, its two most important parameters are $\tau_*$, the characteristic continuum optical depth of the wind as defined in \citet{owo01}, and $R_0$, the radius at which the X-ray emission begins. For He-like triplets, the relative strengths of the components are also derived. While the overall line strengths and the $G$ ratios of the He-like triplets agree well with those found by the first method, the other parameters are ill-constrained; we are  not able to draw more precise conclusions from these  models than we do from our simple fits.

\begin{table*}
\centering
\caption{Results of the Gaussian line fits.  }
\label{rgsfit}
\begin{tabular}{lccccc}
\hline\hline
Parameter  & Mg\,{\sc xi} & Mg\,{\sc xii} & Ne\,{\sc ix} & Ne\,{\sc x} &  Fe\,{\sc xvii} \\
\hline
$v$ (km\,s$^{-1}$) & $-$160$\pm$170 & $-$395$\pm$281 & 804$\pm$255 & $-$73$\pm$155 & $-$139$\pm$118 \\
FWHM (km\,s$^{-1}$)& 600$\pm$631    & 1408$\pm$767   & 2749$\pm$552& 1194$\pm$364  & 1134$\pm$ 343  \\
Global flux ($10^{-5}$\,ph\,cm$^{-2}$\,s$^{-1}$)& 4.37$\pm$0.51 & 3.40$\pm$0.38 & 4.50$\pm$0.56 & 4.13$\pm$0.37 & 4.36$\pm$0.34 \\
Flux of $f$ line  & 1.19$\pm$0.28 &&1.46$\pm$ 0.36 \\
Flux of $i$ lines & 0.28$\pm$0.28 &&0.00$\pm$ 0.11 \\
Flux of $r$ line  & 2.89$\pm$0.32 &&3.04$\pm$ 0.41 \\
\hline
\end{tabular}
\\
\tablefoot{The line fluxes are the observed ones, not corrected for (interstellar) absorption. }
\end{table*}

\section{Discussion}
In this section, we try to interpret, one at a time, the observational results  presented above.

First, \hd\ presents an exceptionally deep (one order of magnitude) and long ($\Delta(\phi)\sim0.1$) X-ray minimum. For comparison, WR21a shows no flat bottom and the X-ray flux changes only by a factor of 4 between apastron and periastron \citep{gos16}. In WR140, the flat minimum lasts only 0.01 in phase and a flux ratio of two is observed between apastron and periastron \citep{cor11}. Only $\eta$\,Car appears to have a very low minimum, with a one order of magnitude decrease in the hard flux  \citep{ham14}, but it lasts only 0.01--0.03 in phase \citep{cor10}. \hd\ thus appears to be an extreme case and the cause for this long and deep minimum needs to be investigated. The question arises  of whether the X-ray emission associated with the wind collision completely disappears at periastron in \hd. When the emission is bright, the high-resolution \xmm\ spectra clearly indicate that the X-ray source is far from the photospheres, hence linked to a distant wind-wind collision. Unfortunately, no high-resolution spectrum is available for the minimum flux phase, so we cannot formally prove that the X-ray source has changed position. However, there is another way to check whether X-rays come from a wind-wind collision. Indeed, O-stars display an intrinsic X-ray emission whose origin lies in embedded shocks distributed throughout the wind. Its intensity is strongly correlated with the bolometric luminosity following $\log(L_{\rm X}/L_{\rm BOL})\sim-7$ with a natural dispersion of a few tenths of dex around that relation \citep[and references therein]{naz09,naz11,rau15}. For \hd, the minimum X-ray flux corresponds to $\log(L_{\rm X}/L_{\rm BOL})\sim-6.85$. It thus appears compatible with the ``canonical'' relation, hence a total extinction of the wind-wind emission seems possible. 

A second question then naturally arises of the reason for that disappearance. In the binaries mentioned above, it is due to an increased absorption, sometimes combined to a collapse of the collision zone onto the companion. In our case, if the minimum was due to a strong extinction, the spectral fitting would reveal a large increase of the absorbing column, but that is not the case. In fact, $N_{\rm H}$ appears quite stable, with only a small-amplitude modulation which cannot explain a flux that decreases by an order of magnitude (Fig. \ref{fitparam}). Another way to decrease the observed flux is an eclipse (see e.g. the case of V444\,Cyg, \citealt{lom15}). \hd\ is indeed an eclipsing system, at least near periastron \citep{ote05}\footnote{\citet{ote05} mentioned an eclipse reference time of 2\,452\,437.660, which corresponds to $\phi$=0.02, a value compatible with the results of the more recent monitoring of Paper I.}. The problem is that this is a grazing eclipse, so it is both short (duration $\Delta(\phi)\sim0.01$ for the flat bottom and $\Delta(\phi)$=0.04 between start and end; see also Fig. \ref{lc}) and shallow ($\Delta(V)\sim$0.2\,mag). Even without considering that the collision zone is larger than the stellar bodies, which would further decrease the impact of an eclipse, such an event thus appears incompatible with the long and deep minimum we observe at X-ray energies. An alternative scenario would be a collapse of the shock. For example, the X-ray emission could lie close to the photosphere of the star with the weakest wind, on the hemisphere facing its companion. In that case, occultation would occur when the star ``turns its back'' towards us, as in CPD--41$^{\circ}$7742 \citep{san05}. In our case, the minimum occurs close to periastron, i.e. when the primary star is in front and the secondary behind, so that the heated stellar hemisphere would have to be on the {\it primary} star. There is, however, no indication from the spectral types and visible spectra analyses that this star has a particularly weak wind (Paper I). The only remaining possibility is a full shock disruption. The wind-wind collision zone would be completely (or nearly completely) destroyed near periastron, erasing the source of X-rays.  This possibility is supported by the estimate of the balance between the two winds (see below), as well as by the disappearance of the H$\alpha$ emission detected in the optical domain at $\phi=0$ (Paper I).

The X-ray light curve of \hd\ presents two other peculiarities: an event with a fast and localized increase near $\phi=0.65$, and a global asymmetry (very long recovery after periastron, steep decrease before it). Regarding the former peculiarity, it is tentative to associate such a flux increase to the line of sight passing through the weaker (hence less absorbing) wind of the secondary star, i.e. the line of sight lies inside the shock cone. However,  the observed absorption is not particularly smaller at that specific phase, and  this could only occur around the second conjunction, and maybe somewhat afterwards in case of important Coriolis deflection (see the case of V444 Cyg, \citealt{lom15}). In \hd, however, the second conjunction occurs at $\phi=0.77$, i.e. $\Delta(\phi)=0.1$ {\it after} the event: the short-term variation detected in the light curve of the long \xmm\ observation, confirmed by the change in normalization factors (Fig. \ref{fitparam}), can thus not be explained in terms of the shock cone crossing the view. A specific monitoring around this phase would be required to better characterize this variation (amplitude and duration of the event), hence to unveil its origin.

\begin{figure}
\includegraphics[width=8.5cm, bb=15 140 590 425, clip]{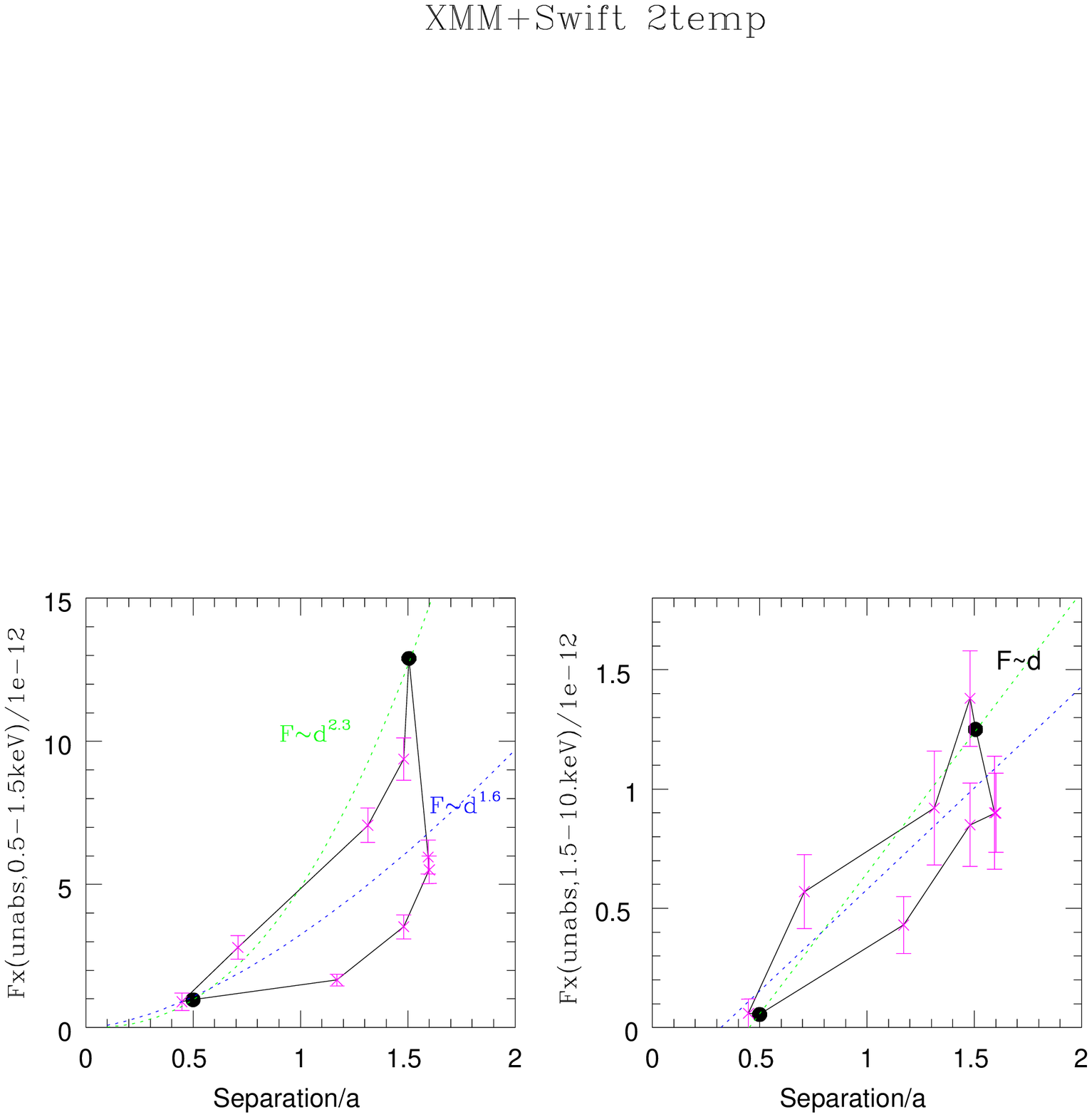}
\caption{Soft (left) and hard (right) fluxes, after correction for the interstellar absorption and in units 10$^{-12}$\,erg\,cm$^{-2}$\,s$^{-1}$, relative to the separation between the components in \hd\ (in units of $a$). \swift\ and \xmm\ observations are shown by magenta crosses and black filled circles, respectively. The best-fit relations are shown by dotted lines (in green when taking errors into account, so \xmm\ data are driving the fit; in blue when all points have the same weight).}
\label{hyst}
\end{figure}

Concerning the latter peculiarity, the recovery after the minimum takes more than a third of the orbit for \hd, a very strong departure from what is seen in $\eta$\,Car, WR140, or WR21a, three cases where shock collapses were also considered \citep[respectively]{cor10,cor11,gos16}. If the minimum is linked to a collapse, \hd\ would present the most extreme example of recovery, although its stars are not as extreme as WRs or LBVs. Furthermore, the decrease towards minimum and increase after it are not symmetric. While this is not unprecedented, in our case it seems correlated to the fact that the two conjunctions occur at $\phi=0.02$ and 0.77 (i.e. the minimum and maximum of the light curve better correlates with the conjunction phases than with the periastron/apastron phases). To investigate this point further, Fig. \ref{hyst} plots the X-ray fluxes as a function of separation. We note that there is a small hysteresis, with a harder flux after apastron (as theoretically expected; see \citealt{pit10}). However, the main result is the detection of a strong correlation with separation. The best-fit relations are $F_{\rm X}\propto d^{2}$ and $F_{\rm X}\propto d$ for the soft and hard bands, respectively (Fig. \ref{hyst}). In adiabatic collisions, a relation $F_X\propto d^{-1}$ is expected -- as is observed for example in WR25 \citep{gosth,pan14} or Cyg OB2 \#9 \citep{naz12} -- thus the collision in \hd\ is certainly {\it not} adiabatic. The alternative scenario, a radiative collision, may lead to a flux modulation which is proportional to separation, as was observed in Cyg OB2 \#8A \citep[Fig. 3 in][and references therein]{caz14}, HD\,152218, or HD\,152248 \citep[Fig. 3 in][]{rau16}. However, in these systems, the hystereses appear more important and the flux variations only reach a factor of about two between minimum and maximum, well below the one order of magnitude observed for \hd, even though these previous cases had much smaller eccentricities than \hd, which could possibly account for the difference. Moreover, we  note that the X-ray luminosity of a radiative wind-wind collision is expected to be proportional to $\dot M v^2$ \citep{ste92}. A  one order of magnitude change in flux would then correspond to a  factor of three change in wind velocity. This would then lead to a one order of magnitude change in temperature   since $kT\propto v^2$. In \hd, however, spectral fits indicate that the temperature does not change much (Sect. 4.2);  indeed, only a small change in hardness is detected between the two sensitive \xmm\ observations (see Sect. 4.1). In contrast, we  recall that for Cyg OB2 \#9, a change in temperature (from 2.5 to 1.9\,keV) was already clearly detected as the wind velocity dropped by 10\% \citep{naz12}. It would therefore be extremely surprising that a large change in wind velocity would go unnoticed in our data. 

\begin{figure*}
\includegraphics[width=9.cm]{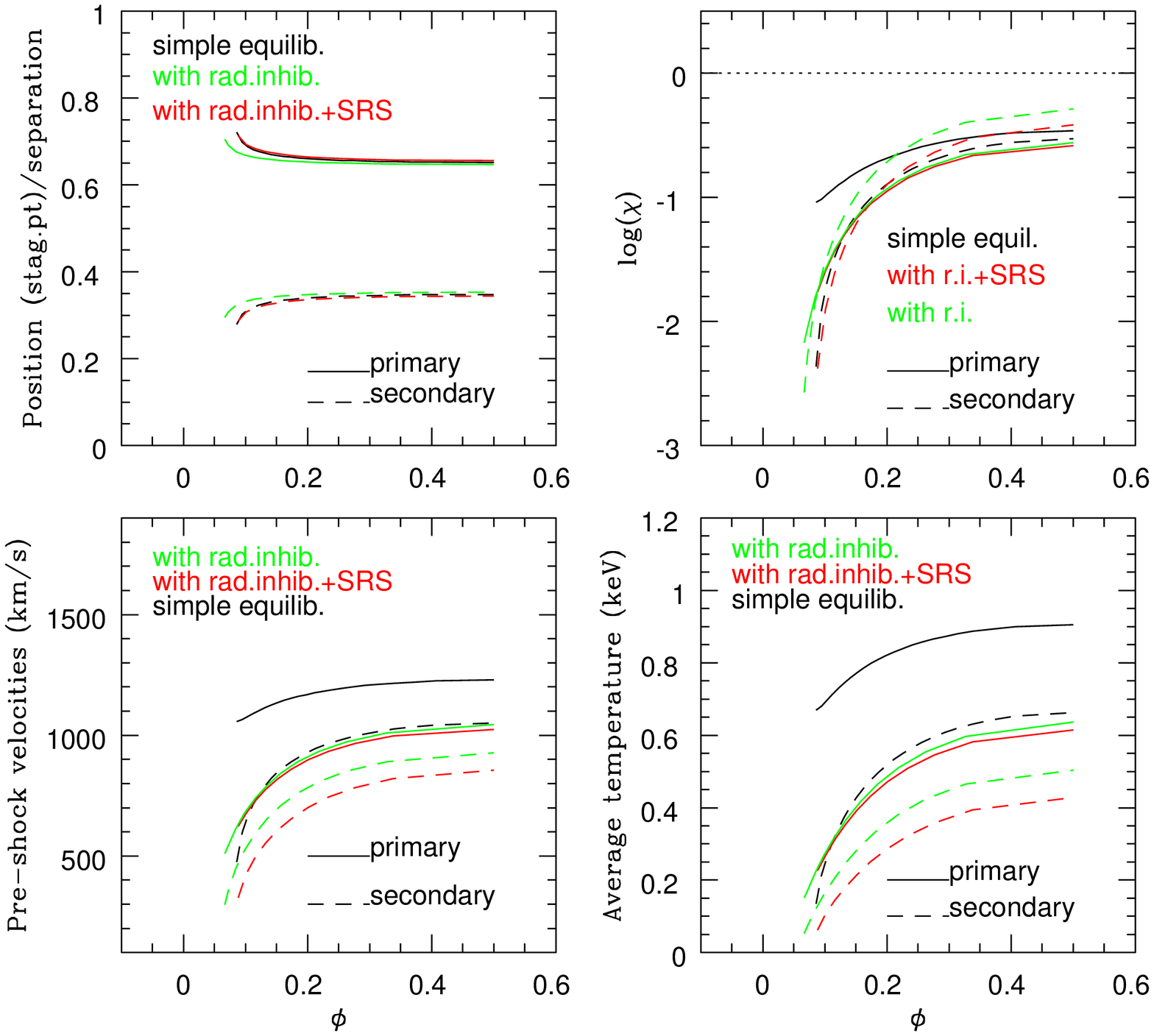}
\includegraphics[width=9.cm]{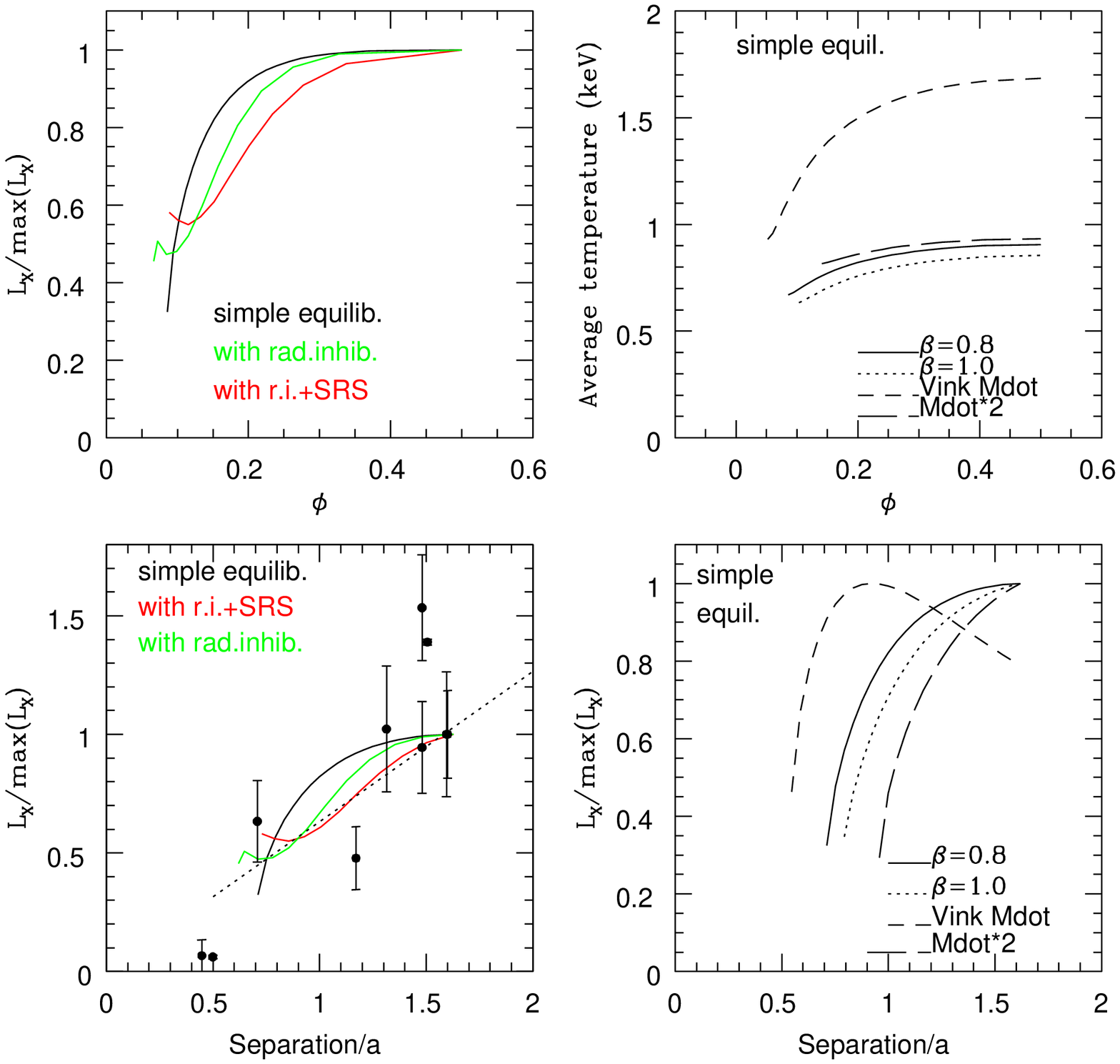}
\caption{{\it Left and Middle panels:} Theoretical evolution with phase of the position of the stagnation point, of the pre-shock velocities, of the cooling parameter $\chi$ (the dotted line in that panel indicating the limit between radiative and adiabatic collisions), of the average post-shock temperatures ($kT\sim 0.6 [v/1000\,{\rm km\,s}^{-1}]^2$), and of the X-ray luminosities (compared to their maximum values). Luminosities are also shown as a function of separation rather than phase. In this panel,  observed fluxes in the hard band, corrected for interstellar absorption and normalized by values closest to apastron, are shown with black dots to facilitate comparison with expectations. In the same panel, the best-fit linear relation to the predicted X-ray luminosities (including radiative inhibition and SRS) is added as a dotted line to highlight the predicted trend with separation. {\it Right panels:} Comparison of the temperatures (top, as a function of phase for the primary wind) and of the normalized X-ray luminosities (bottom, as function of separation) predicted for simple equilibrium using mass-loss rates from Table \ref{param} with velocities following a $\beta=0.8$ or 1.0 law (solid and dotted lines, respectively), using mass-loss rates derived from the recipe of \citet{vin00} with $\beta=0.8$ (short dashed line), or using a mass-loss rate for the primary twice as high as that in Table \ref{param} with $\beta=0.8$ (long dashed line).}
\label{theo}
\end{figure*}

To further investigate this question, we calculated the expected collision parameters in a way similar  to that done for Cyg OB2 \#9 \citep{naz12}. To this end, we considered the stellar and orbital parameters listed in Table \ref{param} (Paper I). We also assume a $\beta$-law (with $\beta$=0.8) for the wind velocities 
and repeated the equilibrium calculation considering radiative inhibition with or without self-regulated shocks \citep[SRS, ][]{par13}. The first two columns of Fig. \ref{theo} show the predicted collision parameters. The equilibrium between the two wind momenta is found to occur at about 65\% of the separation, counting from the primary star, for $\phi>0.1$; no equilibrium between the two winds can be found at and close to periastron. In parallel, the pre-shock velocities decrease with phase, leading to a decrease of at least 40\% in the shocked plasma temperature, while observations suggest a smaller variation (see Sect. 4.2). Finally, the cooling parameter $\chi$, which indicates the nature of the collision \citep[$>1$ if adiabatic, radiative otherwise; ][]{ste92}, suggests that the collision always remains radiative. 

In addition, we calculated the expected X-ray luminosities for each colliding wind using the formalism of \citet{zab11}, and summed them to get the evolution with phase of the total X-ray luminosity associated with the collision. The results are shown in the third column of Fig. \ref{theo}: whatever the case, a maximum at apastron is expected, with a longer recovery after periastron if the effects of radiative inhibition and self-regulated shocks are included. In fact, when plotted relative to separation, theory and observation appear to follow similar trends, especially if the data near $\phi\sim0.65$ are excluded. In this context, we note that these predictions are sensitive to the chosen wind parameters. In particular, if the mass-loss rates of the \citet{vin00} recipe are used, implying a reduction of the primary value by a factor of two, then the collision becomes adiabatic during more than half the orbit, the temperatures are higher, wind equilibrium nearly always occurs, and the X-ray luminosity varies in a very different way, incompatible with observations (see the last column of Fig. \ref{theo}). Increasing the mass-loss rate of the primary by a factor of two yields less dramatic changes, but the collapse occurs even earlier. The relative strengths of the winds are thus well constrained by the X-ray light curve. However, changing the exponent of the velocity law in simple equilibrium cases to $\beta$=1, as is better for supergiants, does  not change  the predictions much (see last column of Fig. \ref{theo}). Nevertheless, the observed X-ray light curve is clearly not symmetric around $\phi=0.5$ as in our simple analytical models. Combined with the too low and variable expected temperatures, this underlines the need for a full hydrodynamic model of the system to secure our understanding of the collision in \hd.

\section{Conclusion}

Using \swift\ and \xmm\ observatories, we performed an X-ray monitoring of the massive binary \hd. It revealed clear evidence of the presence of a wind-wind collision: high flux, presence of a $f$ line combined to the absence of the $i$ lines in triplets of He-like ions, and presence of phase-locked flux variations. The observed changes have three important characteristics: (1) a long and deep minimum occurring at and very close to periastron (and to the time of the optical eclipse), (2) asymmetric decrease/increase in flux, and (3) a short interval of increasing flux after apastron but before conjunction. The spectral analysis revealed no significant change in temperature nor absorption: the flux variations are mostly due to changes in the actual emission strength. Because of the long duration and the quasi-constancy of absorption, (wind) eclipse or occultation effects cannot explain the deep flux decrease. When flux variations are compared to the varying separation, a small hysteresis is detected around trends of the form $F_{\rm X}\propto d$ or $d^2$ for the hard and soft band, respectively. The collision thus does not appear to be adiabatic in nature. Predictions from analytical models of the collision based on the orbital and stellar parameters of the system indeed favour a radiative nature, and seem to explain the observed trend in X-ray luminosity. The agreement is less good for temperatures (too low and variable in models), and some light curve peculiarities (asymmetric character, with an event between apastron and conjunction keeping the X-ray flux rather high) remain unexplained, calling for more sophisticated modelling as well as complementary observations. 

\begin{acknowledgements}
We acknowledge support from the Fonds National de la Recherche Scientifique (Belgium), the Communaut\'e Fran\c caise de Belgique, the PRODEX \xmm\ and Integral contracts, and an ARC grant for concerted research actions financed by the French community of Belgium (Wallonia-Brussels Federation). We thank Norbert Schartel and the ESAC staff, as well as the \swift\ scientists for their kind assistance that made these observations possible. ADS and CDS were used in   preparing this document. 
\end{acknowledgements}

\end{document}